\documentstyle[11pt,iau185,twoside,epsfig]{article}

\begin{document}
\title{RR Lyrae Stars and Anomalous Cepheids in the Draco Dwarf Spheroidal Galaxy}
\author{K. Kinemuchi, H. A. Smith, A. P. LaCluyz\'{e}, C. L. Clark}
\affil{Michigan State University, Physics \& Astronomy Department, East Lansing, MI 48824 USA}
\author{H. C. Harris}
\affil{United States Naval Observatory, P.O. Box 1149, Flagstaff, AZ 86002-1149 USA}
\author{N. Silbermann}
\affil{SIRTF Science Center, California Institute of Technology, Mail Code 220-6, 1200 East California Boulevard, Pasadena, CA 91125 USA}
\author{L. A. Snyder}
\affil{DePauw University, Physics \& Astronomy Department, 602 South College Avenue, Greencastle, IN 46135 USA}

\begin{abstract}
We present new results on RR Lyrae stars and anomalous Cepheids in the Draco
dwarf spheroidal galaxy.  We have increased the number of double-mode RR Lyrae
stars 
and found three new anomalous Cepheids.  With period-magnitude and
period-amplitude diagrams, we discuss the Oosterhoff classification of Draco.
Contradictory results were found in that Draco appears to contain both
Oosterhoff I and II type RR Lyrae populations.
\end{abstract}


\section{Introduction}
The Draco dwarf spheriodal galaxy was first extensively studied by Baade \& 
Swope (1961), who identified 260 variable stars in their survey, determining 
periods and photographic B light curves of 138 variables.  
Goranskij (1982) and Nemec (1985) reanalyzed the Baade \& Swope observations,
finding 10 double-mode RR Lyraes (RRd).  In this current survey of Draco,
we present our photometric results of 268 RR Lyraes and 8 anomalous Cepheids.

The Draco dwarf spheroidal galaxy is a metal poor system with an 
$\langle {\rm [Fe/H]} \rangle$
$\sim -2.0$ (Mateo, 1998).  Draco has a well populated red horizontal
branch (HB).  However, compared to a metal poor globular cluster such as M15,
Draco has a sparsely populated blue horizontal branch.

\section{Observations and reduction}
The data were obtained from two sources: the 1.0m telescope at the United 
States Naval Observatory (USNO) - Flagstaff station and the 2.3m telescope
at the Wyoming Infrared Observatory (WIRO).  The observations spanned
over two years, from June 1993 to July 1995.  There were between 40 to 
60 observations in $V$ and $I$.

The images were reduced using the standard IRAF CCD reduction packages
while the photometry was done with the stand alone versions of 
DAOPHOT II and ALLFRAME (Stetson, 1987, 1994).  To find the periods,
two methods were utilized: a date compensated discrete Fourier 
transform supplemented by phase dispersion minimization 
(Stellingwerf, 1978).

\section{Results}
We have increased the number of RRd stars from the original 10 to 
at least 26.  Prior to this work, five anomalous Cepheids were known 
to be in the Draco dwarf galaxy, and to this number we have added three more.

In the period-magnitude diagram, Fig.\,1, a clear separation between 
the ab and cd type RR Lyraes is shown.  Nemec (1988) proposed that 
two period-luminosity (P-L) relationships exist depending on the pulsational 
mode (fundamental or first overtone) of anomalous Cepheids for five dwarf
spheroidal galaxies.  Three of the newly discovered anomalous Cepheids of
Draco fall close to Nemec's fundamental mode P-L relation, but it is not 
clear if the Draco Cepheids fall along two distinct P-L relations. 
\begin{figure}[hp]
\epsfxsize = 6.5 cm
\centerline{\epsfbox{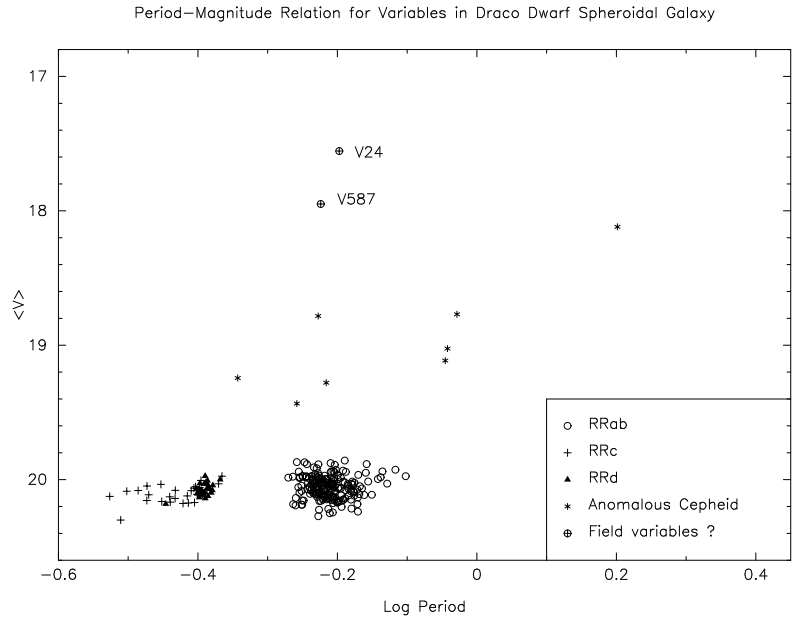}}
\caption{Period-magnitude diagram for RR Lyraes and anomalous Cepheids in
Draco.  Included are two possible field variable stars, labeled V24 and V587.}
\end{figure}

A period-amplitude diagram for Draco RR Lyraes is presented in Figure 2.
Here we have overlaid the Oosterhoff I (M3) and Oosterhoff II ($\omega$ 
Centauri) trend lines for RRab stars (Clement \& Rowe, 2000).  The Draco 
RRab stars appear to mostly fall along or are close to the Oosterhoff I 
line.  In comparison, galactic globular clusters that are more metal poor
than [Fe/H] = -1.7 are classified as Oosterhoff II.
\begin{figure}[hp]
\epsfxsize = 6.5 cm
\centerline{\epsfbox{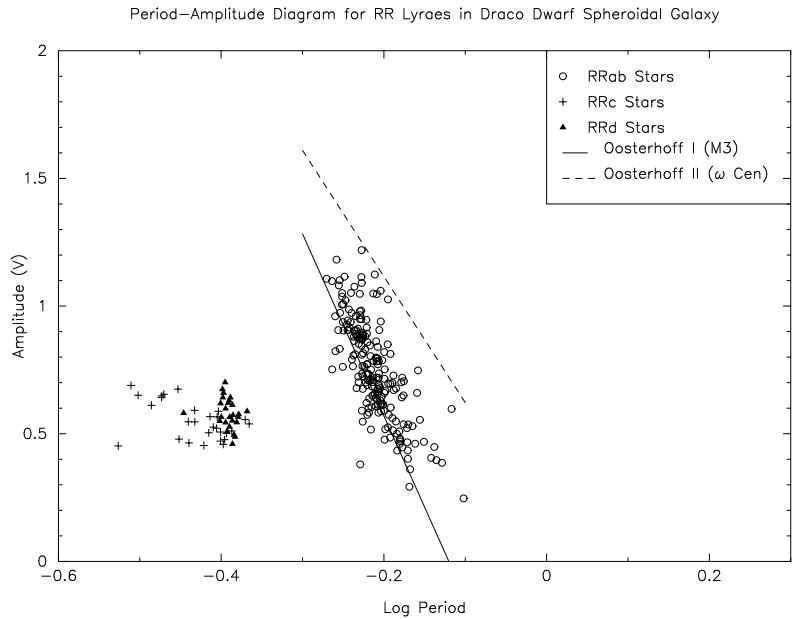}}
\caption{Period-amplitude diagram for Draco RR Lyraes.}
\end{figure}

Lee et al.\ (1990) provide a plausible explanation for
this contradiction in Oosterhoff classification in Draco.  RR Lyraes
in Oosterhoff II globular clusters are originally on the blue horizontal
branch and evolve away from their zero-age horizontal branch positions.
As they evolve, the stars become brighter and redder.  According to Ritter's
relation, as the star gets brighter, the period becomes longer.  
Since Draco has a sparsely populated blue horizontal branch, this is
not the best description for this system.  If the Draco RRab stars 
are not as evolved as those in a metal poor globular cluster, then
these stars are fainter, have shorter periods, and hence fall closer
to the Oosterhoff I trend line.
  
However, the RRc and RRd stars of Draco have periods similar to those
found in Oosterhoff II globular clusters.  Only one RRd star
was found to be similar to Oosterhoff I globular cluster RRd stars.  Some
c-type RR Lyraes are also like Oosterhoff I stars, but a majority of them
are Oosterhoff II. Detailed modeling for different evolutionary 
scenarios is needed to see how difficult it is to explain the different 
Oosterhoff properties of RRab and RRcd stars.

\section{Future work}
In this paper we presented our results from $V$ band photometry only.  $V$ and
$I$ colors are being determined.  We will also be investigating the prevalence
of the Blazhko effect among the RR Lyraes in Draco and the radial 
distributions of the variable stars.  

\acknowledgments This work has been supported in part by the National Science
Foundation under grants AST-9528080 and AST-9986943.  L.A. Snyder was partially
supported by the NSF REU grant 991221.

\section*{Discussion}

{\it D. Laney~:} Could the peculiar distribution of RR Lyrae stars with 
respect to Oosterhoff I and II be explained by an age distribution 
different from that found in typical Oosterhoff I and II globular 
clusters? \\ [0.2cm]
{\it K. Kinemuchi~:} Grillmair et al. (1998) discuss a lack of multiple 
main sequence turnoffs; thus no multiple star formation episodes.  From their
analysis, there is no large age distribution in Draco.  \\ [0.4cm]
{\it G. Kov\'{a}cs~:} The P-A diagram may not necessarily be the best way
to estimate cluster metallicity.  For RRab stars, there is a much better
correlation with P and $\phi_{31}$.  Also, amplitudes might be affected in 
the case of crowded field photometry.  Concerning the latter effect, do 
you have an estimation on the severity of crowding in the case of your data 
set? \\ [0.2cm]
{\it K. Kinemuchi~:} Draco is not a very compact object, so crowding does
not usually become an issue.  We do have problems with near neighbors, but 
almost all of that is taken care of with our photometry programs.  We have not 
yet looked into the $\phi_{31}$ values of our RRab light curves.


\begin{references}
\reference Baade, W. \& Swope, H. 1961, \aj, 66, 300
\reference Clement, C. \& Rowe, J. 2000, \aj, 120, 2579
\reference Goranskij, V. P. 1982, Astronomicheskij Tsirkulyar, 1216, 5
\reference Grillmair, C. J. et al. 1998, \aj, 115, 144
\reference Lee, Y. W., Demarque, P. \& Zinn, R. 1990, \apj, 350, 155
\reference Mateo, M. 1998, \araa, 36, 435
\reference Nemec, J. 1985, \aj, 90, 204
\reference Nemec, J., Mendes de Oliveira, C., \& Wehlau, A. 1988, in 
ASP Conf. Ser, Vol. 4, The 
Extragalactic Distance Scale,
eds. S. van den Bergh \& C. Pritchet, (San Francisco ASP), 180
\reference Stellingwerf, R. F. 1978, \apj, 224, 953
\reference Stetson, P. B. 1987, \pasp, 99, 191
\reference Stetson, P. B. 1994, \pasp, 106, 250
\end{references}
\end{document}